\documentstyle[12pt]{article}
\def\V{{\cal V}}
\def\H{{\cal H}}
\def\C{{\cal C}}
\def\K{{\cal K}}
\def\L{{\cal L}}
\def\ka{{\kappa}}
\newcommand{\be}{\begin{eqnarray}}
\newcommand{\en}{\end{eqnarray}}

\begin{document}
\begin{titlepage}
\begin{flushright}
EFI 94-39

MPI-Ph/94-52
\end{flushright}

\begin{center}
\vskip 0.3truein

{\bf {Analytic Structure of Amplitudes}}
\vskip0.10truein

{\bf {in Gauge Theories with Confinement}}
\footnote{Based on an invited talk presented at the Conference on New
   Problems in the General Theory of Fields and Particles  of the
   {\it International Congress of Mathematical Physics}, La Sorbonne,
   Paris, July 1994.}
\vskip0.5truein

{Reinhard Oehme}
\vskip0.2truein

{\it Enrico Fermi Institute and Department of Physics}

{\it University of Chicago, Chicago, Illinois, 60637, USA}
\footnote{Permanent Address}

{\it and}

{\it Max-Planck-Institut f\"{u}r Physik}

{\it - Werner-Heisenberg-Institut -}

{\it 80805 Munich, Germany}
\end{center}
\vskip0.2truein
\centerline{\bf Abstract}
\vskip0.13truein

For gauge theories with confinement, the analytic
structure of amplitudes is explored.
It is shown that the analytic properties of {\it physical}
amplitudes are the same as those obtained on the
basis of an effective theory involving only the
composite, physical fields. The corresponding proofs
of dispersion relations remain valid.
Anomalous thresholds are considered. They are
related to the composite structure of particles.
It is shown, that there are no such thresholds in
physical amplitudes which are associated with confined
constituents, like quarks and gluons in QCD.
{\it Unphysical} amplitudes are considered briefly, using
propagator functions as an example. For
general, covariant, linear gauges, it is
shown that these functions must have singularities
at finite, real points, which may be associated with
confined states.

\end{titlepage}
\newpage
\baselineskip 20 pt
\pagestyle{plain}

\centerline{\bf I. INTRODUCTION}
\vskip0.2truein
The analytic structure of amplitudes is
of considerable importance in all quantum field theories,
from a physical as well as a conceptional point of view. It
has been studied extensively over many years, but mainly for
field theories with a state space of definite metric, and in
situations where the interpolating Heisenberg fields are
closely related to the observable excitations of the theory.
Within a relativistic framework of this type, the
commutativity or anti-commutativity of the Heisenberg
fields, at space-like separations, gives rise to tubes
(wedges) of holomorphy for retarded and advanced amplitudes,
which are Fourier transforms of tempered distributions.
Lower bounds for the spectrum of eigenstates of the energy
momentum operator provide real domains where these
amplitudes coincide at least in the sense of distributions.
One can then use the Edge of the Wedge theorem \cite{BOT}
to show that
there exists an analytic function which is holomorphic in
the union of the wedges and a complex neighborhood of the
common real domain, and which coincides with the advanced
and retarded amplitudes where they are defined. Then the
envelope of holomorphy \cite{BRO, BOT} of this
initial region of analyticity
gives the largest domain of holomorphy obtainable from the
general and rather limited input. Further extensions require
more exhaustive use of unitarity \cite{MAT}, which is often rather
difficult. Although the theory of functions of several
complex variables is the natural framework for the
discussion of the analytic structure of amplitudes, for
special cases, like those involving one complex four-vector,
more conventional methods, like differential equations and
distribution theory, can be used in order to obtain the
region of holomorphy \cite{JLD, BOG}. Many more technical details
are involved in the derivation of analytic properties
and dispersion relations \cite{DSF, SYM},
\cite{BOG, BOT, LEH, OTG}, but the envelopes mentioned
above are at the center of the problem.

It is the purpose of this paper to discuss essential
aspects of the analytic structure of amplitudes for gauge
theories, which require indefinite metric in a covariant
formulation, and for which the physical spectrum is not
directly related to the original Heisenberg fields.
Rather, these fields
correspond to unphysical, confined excitations in the state
space of indefinite metric. We will be mainly concerned with
physical amplitudes, corresponding to hadronic
amplitudes for QCD. We will often use the language of QCD.
As examples of unphysical
Green's functions, we consider the structure functions of
the gluon and quark propagators. In all covariant,
linear gauges, we show that these generally cannot
be entire functions, but must have singularities which can
be related to unphysical states. A preliminary account
of some of our results may be found in \cite{OPN}.

In the framework of hadronic field theory with positive
definite metric, the derivation of analytic properties, and
of corresponding dispersion relations, is on a quite
rigorous basis, and uses only very general aspects of the
theory. For gauge theories with confinement however, the
derivation of dispersion relations requires several
assumptions, which may not have been proven rigorously
in the non-perturbative framework required in the presence of
confinement. We will discuss these assumptions in the
following sections. They mainly concern the definition
of confinement on the basis of the BRST-algebra,
and the construction of composite hadron fields as
products of unphysical Heisenberg fields.

In order to provide for the input for proofs of dispersion
relations and other analyticity properties of physical
amplitudes, we discuss in the following paragraphs certain
results for non-perturbative gauge theories. For some of these
results, we can refer to
the literature for detailed proofs, but we have to explore
their relevance for the derivation of analytic properties
and for the identification of singularities.
It is not our aim, to provide reviews of several
aspects of non-perturbative gauge theories. What we
need is to show that there is a mathematically well
defined formulation of confinement, which we use in
order to derive the spectral conditions, and a construction
of local, composite fields for hadrons, from which we
obtain the initial domain of holomorphy. We also have to
exhibit the assumptions made within this framework.

For the purpose of writing Fourier representations for
hadronic amplitudes, we must consider the
construction of local, BRST-invariant, composite
Heisenberg fields corresponding to hadrons.
This problem requires a discussion of operator
products \cite{WZC, WLS}
of elementary, confined fields in a
non-perturbative framework.

Since we need a manifestly covariant formulation of the
theory, we consider linear, covariant gauges within the
framework of the BRST-algebra \cite{BRS}. Assuming the existence and
the completeness \cite{SPI} of a nilpotent
BRST-operator $Q$ in the state
space $\V$ of indefinite metric, we define an invariant
physical state space $\H$ as a cohomology of $Q$. As a
consequence of completeness, which implies that all neutral
(zero norm) states satisfying $Q\Psi = 0$ are of the form
$\Psi = Q\Phi, \Phi \in \V $, the
space $\H$ has (positive) definite metric
\cite{KOJ, ROC}.

We assume that there are hadronic states in the theory, and
take confinement to mean that, in a collision of hadrons,
only hadrons are produced. Within the BRST-formalism, this
implies that only hadron states appear as physical states in
$\H$. At least at zero temperature, transverse gluons and
quarks are confined for dynamical reasons, forming
non-singlet representations of the BRST-algebra
in combination with other unphysical fields. In the
decomposition of an inner product of physical states, there
appear then only hadron states. The same is true for the
decomposition of physical matrix elements of products of
BRST-invariant operator fields, because these operators map
physical states into other physical states. In this way we
find that the fundamental {\it spectral conditions} for
hadronic amplitudes are the same as in the effective hadronic
field theory. The absorptive thresholds are due only to
hadronic states.

There is, however, another category of singularities, which
is related to the structure of particles as a composite
system of other particles. These are the so-called anomalous
thresholds or structure singularities. They were encountered
in the process of constructing examples for the limitations
of proofs for dispersion representations
\cite{RAN, NAM}. These limitations
are related to anomalous thresholds corresponding to the
structure of a given hadron as a composite system of
non-existing particles, which are not excluded by simple
spectral conditions. But physical anomalous thresholds
\cite{KSW, NAM, RHF} are
very common in hadronic amplitudes: the deuteron as a
np-system, $\Lambda$ and $\Sigma$ hyperons as KN systems,
etc. In theories like QCD, the important question is, whether
there are structure singularities of hadronic amplitudes
which are related to the quark-gluon structure of hadrons.
We show that this is {\it not} the case. Independent of
perturbation theory, we describe how anomalous thresholds
are due to poles or absorptive thresholds in crossed
channels of other hadronic amplitudes, which are related to
the one under consideration by analytic continuation into an
appropriate lower Riemann sheet \cite{RHF}.
Since, as explained above,
we have no absorptive singularities in hadronic amplitudes
which are associated with quarks and gluons, we also have no
corresponding anomalous thresholds.

For form factors of
hadrons, which may be considered as loosely bound systems of
of heavy quarks, there are interesting consequences
stemming from the absence of anomalous thresholds
associated with the quark-gluon structure.
In a constituent picture, these hadrons can be
described by a Schr\"odinger wave function with a long range
due to the small binding energy. But in QCD, in contrast to
the situation for the deuteron, for example, there are no
anomalous thresholds associated with the spread-out quark
structure. However, there is no problem in obtaining a large
mean square radius with an appropriate form of the
discontinuities associated with hadronic branch lines. In
addition, there may be {\it hadronic} anomalous thresholds
which are relevant.

Finally, we consider the analytic properties and the
singularity structure of unphysical (colored) amplitudes. It
is sufficient to discuss two-point functions as examples.
The structure functions of quark and transverse
gluon propagators are analytic in the $k^2$-plane,
with cuts along the positive real axis. This is a direct
consequence of Lorentz covariance and spectral conditions.
In previous papers \cite{WZS},
we have derived the asymptotic behavior
of these functions for $k^2\to\infty$ in all directions of
the complex $k^2$-plane, and for general, linear,
covariant gauges \cite{OWG}. With asymptotic freedom, they
vanish in these limits. Hence the structure functions cannot
be non-trivial, entire functions. They must have
singularities on the positive, real $k^2$-axis, which should
be associated with appropriate unphysical (colored) states
in the general state space $\V$ with indefinite metric.
These states are not elements of the physical state space,
but form non-singlet representations of the BRST-algebra.

\vskip0.6truein
\setcounter{equation}{0}

\centerline{\bf {II. CONFINEMENT}}
\vskip0.2truein

In this section, we briefly define the general framework for
the later discussion of amplitudes and their
analytic structure.

We consider quantum chromodynamics and similar theories.
Since it is essential to have a manifestly
Lorentz-covariant formulation, we use covariant gauges as
defined by a gauge fixing term
\be
\L_{GF} = B \cdot (\partial_\mu A^\mu ) +
\frac{\alpha}{2} B \cdot B ~ ,
\label {2.1}
\en
where $B$ is the
Nakanishi-Lautrup auxiliary field, and $\alpha$ is a real
parameter. The theory is defined in a vector space $\V$ with
indefinite metric. We assume that the constrained system is
quantized in accordance with the BRST-algebra \cite{BRS} :
\be
Q^2 = 0 , ~~~  i[Q_c, Q] = Q ~,
\label {2.2}
\en
where $Q$ is the BRST-operator, and $Q_c$ the ghost-number
operator. On the basis of this algebra, we define the
subspaces
\be
kerQ = \{\Psi : ~Q\Psi = 0, ~ \Psi \in \V \} ~ ,
\label {2.31}
\en
\be
imQ = \{\Psi : ~ \Psi = Q\Phi, ~ \Phi \in \V \} ~ ,
\label {2.32}
\en
where $imQ \perp kerQ$,  with respect to the
indefinite inner product $(\Psi, \Phi)$. We can write
\be
kerQ = \V_p \oplus imQ ~ ,
\label {2.4}
\en
and define the BRST-cohomology
\be
\H ~=~ \frac{kerQ}{imQ}
\label {2.5}
\en
as a covariant space
of equivalence classes, which is isomorphic to $\V_p$
\cite{KOJ}. We are
interested in zero ghost number, and hence ignore the
grading due to the ghost number operator $Q_c$. In order to
use $\H$ as a physical state space, it must have definite
metric, which we can choose to be positive. This is not
assured, a priori, but requires the assumption of
``completeness'' of the BRST-operator $Q$, which implies that all
neutral (zero norm) states in $kerQ$ are contained in $imQ$
\cite{SPI, KOJ}. Then
$\V_p$ and hence $\H$ must be definite, because every space
with indefinite metric contains neutral states. With
completeness, we have $imQ = (kerQ)^{\perp}$ ,
and hence $imQ$ is the isotropic
part of $kerQ$. It is not enough for the definiteness
of $\H$ to have ghost number zero, since `singlet pair'
representations, containing equivalent numbers of ghosts and
anti-ghosts, must also be eliminated. In view of the inner
product for eigenstates of $iQ_c$ :
\be
(\Psi_{N_c} , \Psi_{N'_c} ) ~ = ~ \delta_{-N_c, N'_c} ~,
\label {2.6}
\en
they would give
rise to an indefinite metric in $\V_p$, and hence to neutral
states.

There are arguments for the absence of singlet pairs in
the dense subspace generated by the field operators. But we
are dealing with a space of indefinite metric, so that the
extension to the full space $\V$ is delicate
\cite{SKO, SKU, SNA}. Completeness
can be proven explicitly in certain string theories,
however these are more simple structures than
four-dimensional gauge theories. In any case, without
completeness, we cannot get a physical subspace with
definite norm, and a consistent formulation of the theory
would seem to be impossible.

Given completeness, physical states $\Psi_p$ are BRST-singlets
with $Q\Psi_p = 0$, positive norm and ghost number zero.
Unphysical states form quartet representations of the
BRST-algebra \cite{KOJ}:
\be
{\Psi_{N_c} ~~~ with ~~~ Q\Psi_{N_c} \neq 0 ~,}
\en
\be
{\Xi_{N_c + 1} = Q\Psi_{N_c} ~,}
\en
\be
{\Psi_{-N_c - 1} ~~~ with ~~~ Q\Psi_{-N_c - 1} \neq 0
{}~~~ and ~~~ (\Xi_{N_c + 1} , \Psi_{-N_c - 1} ) \neq 0 ~,}
\en
\be
{\Xi_{-N_c} = Q\Psi_{-N_c - 1} ~~~ and
{}~~~ (\Xi_{-N_c} , \Psi_{N_c} ) \neq 0 ~. }
\label {2.7}
\en
The states $\Psi_{-N_c - 1}$  and $\Xi_{-N_c}$  are
implied by the non-degeneracy of $\V$, and the inner product
in Eq.(\ref {2.6}).

In weak coupling perturbative theory, the state space $\H$
consists of quarks and transverse gluons. Ghosts,
longitudinal- and timelike gluons form quartet
representations, and are unphysical. They are confined in a
kinematical fashion.

In a general non-Abelian gauge theory like QCD, we can have
asymptotic freedom, and we expect that all colored states
are confined, provided the number of matter fields is
limited. In the language of QCD, this implies that quarks
and transvers gluons, at zero temperature, are not elements
of the physical state space $\H$, which then contains only
hadrons as colorless, composite systems \cite{SKO, NIC, ROC}.
Under these circumstances, only hadrons can be produced in
a collision of hadrons. This algebraic notion of confinement
should be compatible with more intuitive pictures of quark
confinement, and with two-dimensional models.
However, for gluons,
two-dimensional models are useless, because there are no
transvers gluons in two dimensions. If the number of flavors
in QCD is limited, we can give arguments that
gluons are not elements of the physical subspace
\cite{ROC, NIC}. These
arguments are based upon superconvergence relations
satisfied by the structure function of the gluon propagator,
which provide a connection between short- and long distance
properties of the theory \cite{WZS, OWG, ROS}.

In our discussion of analytic properties of hadronic
amplitudes, we take it for granted that
confinement is realized in the sense that
the physical state space $\H$ contains only hadronic states.
Quarks and gluons are not BRST-singlets. Together with other
unphysical states, they form quartet representations of the
BRST-algebra and remain unobservable.
\vskip0.6truein

\centerline{\bf III. LOCAL HADRONIC FIELDS}
\vskip0.2truein

Having defined the general state space of the gauge theory
with confinement, we now turn to the problem of constructing
local Heisenberg operators, which can be used as
interpolating fields in amplitudes describing reactions
between physical particles (hadrons),
and in form factors of hadrons. The
construction of composite operators, and of operator
product expansions, has been discussed extensively in the
literature \cite{WZC, WLS}.
The relatively new aspects in our case are the
state space of indefinite metric, and the fact that the
constituents are unobservable. In addition, in view of
confinement, we cannot use perturbation theory methods, and
consequently some assumptions are needed for the
non-perturbative construction of composite fields.

In the following, we discuss the problem with the help of a
generic example. We consider the construction of a meson
field $B(x)$ in terms of fundamental fields $\psi (x)$
and $\overline{\psi}(x)$,
ignoring all inessential aspects like indices etc.. Hence, our
formulae in the following are rather symbolic. The field
$B(x)$ must be local and BRST-invariant, so that
$B(x) \Psi $ is a representative of a physical
physical state, provided $\Psi $ is one.

Let us first consider the product
\be
B(x,\xi) = \psi (x + \xi )  \overline {\psi} (x - \xi ) ~.
\label {3.1}
\en
With $| k, M \rangle $ being a one particle hadron state
with $k^2 = M^2 $, we
assume that this state exists as a composite system, so that
we have a non-vanishing matrix element
\be
\langle 0 |  B(x,\xi) | {k, M}  \rangle  \neq 0 ~,
\label {3.2}
\en
where $| 0 \rangle $ denotes the vacuum state, and where the inner
product involved in Eq.(\ref {3.2}) is the indefinite product
defined in the general state space $\V$. We now define a
Poincar\'e covariant, local operator by the weak limit
\be
B_F (x) = \lim_{\xi \rightarrow 0} \frac {\psi (x + \xi)
\overline{\psi} (x - \xi)}{F(\xi)} ~.
\label {3.3}
\en
We may consider a space-like approach with
$\xi^2  < 0$ , but this is not
essential. The invariant function $F(\xi)$ is only of relevance
as far as its singularity for $\xi \rightarrow 0$ is concerned.
It is the
purpose of $F(\xi)$ to compensate the singularity of the
operator product. Writing
\be
F(\xi) = (\Psi , B(0,\xi) \Phi )~ , ~~~
\Psi , \Phi \in \V ~,
\label {3.4}
\en
we want to choose these states so that they
belong to a class $\K_{max}$ , for which the matrix element
(\ref {3.4}) is most singular, assuming that such most singular
matrix elements exist \cite{WZK}.
Possible oscillations in the limit
(\ref {3.3}) may require the choice of an appropriate sequence
$\{ \xi_n \} $ in the approach to $\xi = 0 $.
By construction, the operator $B_F (x) $
is local with respect to the constituent
fields $\psi (x)$ and $\overline{\psi} (x) $,
and with respect to itself.

In view of the requirement (\ref {3.2}), we have
\be
\langle 0 | B(x, \xi) | k , M \rangle  =
e^{-ik\cdot x}\langle 0 | B(0, \xi) | k , M \rangle  =
e^{-ik\cdot x} F_k (\xi) ~,
\label {3.5}
\en
with $F_k (\xi ) \neq 0$.
Then the operator field
\be
B(x) ~= ~\lim_{\xi \rightarrow 0} \frac{B(x,\xi)}{F_k (\xi)}
\label{3.6}
\en
has a finite matrix element. We may assume that
$F_k (\xi) \in \K_{max} $, so that $B(x)$ appears
as the leading term in the general operator product
expansion of $B(x, \xi )$. However, by construction,
the field $B_F (x)$ should be a BRST-invariant
operator. Since we are dealing only with matrix
elements of $B_F (x) $  with respect to states in
the physical state space $\H$, it is sufficient to
assume that $F_k (\xi ) \in \K'_{max} $, where
$\K'_{max} $ refers only to states in $\H$. We then
use the field $B(x)$ in Eq.(\ref{3.6}) as the Heisenberg
field interpolating between the corresponding
asymptotic states. We introduce asymptotic fields
$B_{in} (x)$ using the free retarded function
$\Delta_R (x - x', M )$ in the Yang-Feldman representation,
and apply the conventional LSZ-reduction formalism
\cite{LSZ} in order to obtain representations of
physical amplitudes in terms of products of $B(x)$
fields. An example would be the  S-matrix element
\be
\langle k',M; p',M |S | k,M; p,M \rangle &=& \frac {1}{(2\pi )^3}
\int \int
d^4 x' d^4 x \exp [ik'x'-ikx]    \cr
&\times& K_{x'} K_x \langle p', M | T B(x') \overline{B} (x) |
p, M \rangle ~ ,
\label{3.7}
\en
or corresponding expressions in terms of retarded or
advanced products. We can also make further reductions
as required for the proofs of dispersion representations.
The reduction method is used here only within the space
$\H$ with definite metric and in a framework without infrared problems.

In four dimensions, the existence of operator expansions, and
of composite operators like $B(x)$, can be proven within the
framework of renormalized perturbation theory
\cite{WZK}, but not yet in
the general theory, as required for our purpose. Hence we
have to make the technical assumptions described above. In
many lower dimensional theories, operator product expansions
are known to exist independent of perturbation theory. They
are expected to be a general property of local field
theories.

The construction of interpolating, hadronic Heisenberg
operators, like $B(x)$ in our example, is of course not unique.
But the different possibilities belong to the same Borchers
class \cite{BOC}.
Different fields in a given class, which have the
same asymptotic fields, define the same S-matrix. It can be
shown that locality is a transitive property: two fields,
which commute with a given local field, are local themselves
and with respect to each other. We have equivalence classes
of local fields. Whatever the construction of a composite
operator like $B(x)$, the resulting fields all are local with
respect to the fundamental fields, and hence belong to the
same Borchers class.
Although we use Borchers theorem here essentially only in the
physical subspace, it can be generalized to spaces with
indefinite metric. The proof involves the equivalence of
weak local commutativity and CPT-invariance, as well as the
Edge of the Wedge Theorem. Introducing appropriate rules for
the transformation of ghost field under CPT, we can define
an anti-unitary CPT-operator in the state space $\V$.
Together with the postulates of indefinite metric field theory, we
then get equivalence classes of local Heisenberg fields in
gauge theories like QCD.

The construction of composite hadron fields, as described
above, can be generalized to other products of fundamental
fields which form color singlets.
\vskip0.6truein
\centerline{\bf IV. SPECTRAL CONDITIONS}
\vskip0.2truein

In the previous section, we have described how we can obtain
representations of hadronic amplitudes in QCD in terms of
local Heisenberg fields, which are BRST-invariant and
interpolate between asymptotic states of non-interaction
hadrons. While the local commutativity of the hadron fields
implies support properties of Fourier representations which
give rise to tubes (wedges) as regions of holomorphy, the
spectral conditions define the real domain where retarded
and advanced amplitudes coincide, generally in the sense of
distributions. Given completeness of the BRST-operator $Q$,
it is convenient for our further discussion to introduce a
self-adjoint involution $\C$ in $\V$, which converts the
indefinite inner product into a definite product denoted by
\be
(\Psi , \Phi )_{\C} = (\Psi , \C \Phi ) ~,
\label{4.1}
\en
where $\C^\dagger = \C$ and ${\C}^2 = 1 $
\cite{KOJ, SPI, ROC}.
With respect to the definite product, we obtain
a decomposition of $\V$ in the form
\be
\V = \V_p \oplus imQ \oplus imQ^*  ~,
\label{4.2}
\en
where $Q^* = \C Q \C$ and $Q^{*2} = 0 $.
With completeness of $Q$, the subspace $\V_p$
has (positive) definite metric, while $imQ$ and $imQ^*$
contain conjugate pairs of neutral (zero norm) states, so
that for every $\Psi \in imQ $, there is a $\Psi' \in
imQ^* $  with $(\Psi ,\Psi') \neq 0 $, while both
states are orthogonal to $\V_p$. Here and in the following we
ignore the grading due to the ghost number operator, since we
are mainly interested in $N_c = 0$. It is convenient to introduce
a matrix notation, writing a vector $\Psi \in \V $ in the form
\be
\Psi  = \left ( \begin{array}{c} \psi_1 \\ \psi_2 \\
\psi_3 \end{array} \right) ~,
\label{4.3}
\en
with components referring to the subspaces $\V_p$, $imQ$
and $imQ^*$ of the decomposition (\ref{4.2}). Then
\be
\C = \left( \begin{array}{clcr}
1 & 0 & 0  \\
0 & 0 & 1  \\
0 & 1 & 0
\end{array}  \right) ~,
\label{4.4}
\en
and the inner product is given by
\be
(\Psi ,\Phi ) = (\Psi, \C \Phi )_{\C} =
\psi_1^* \phi_1 + \psi_2^* \phi_3 + \psi_3^* \phi_2 ~.
\label{4.5}
\en
The BRST-operator can be written as
\be
Q = \left( \begin{array}{clcr}
0 & 0 & 0  \\
0 & 0 & q  \\
0 & 0 & 0
\end{array} \right)   ,
\label{4.6}
\en
with $q$ being an invertible suboperator \cite{ROC}.
$\Psi \in kerQ $ is characterized by $\psi_3 = 0 $,
and a representative of a physical state $\Psi \in \H$ by
$\psi_3 = 0$, $\psi_1 \neq 0$.
Hence, for $\Psi , \Phi  \in \H$, we have $(\Psi , \Phi ) =
\psi_1^* \phi_1 $ in Eq.(\ref{4.5}).
Since $\V_p$ is a non-degenerate subspace, we can
introduce a projection operator $P(\V_p)$ with
$P^2 = P^\dagger = P $.

For the purpose of spectral condition, we are interested in the
decomposition of an inner product with respect to a complete
set $\{ \Psi_n \}$ of states in $\V$,
in particular eigenstates of the
energy momentum operator.
For $\Psi , \Phi \in \V$, we have then
\be
(\Psi , \Phi ) = \sum_n (\Psi , \Psi_n )(\Psi_n ,\Phi ) ~.
\label{4.7}
\en
But if we consider only states $\Psi , \Phi \in kerQ$ ,
we obtain
$(\Psi ,\Phi) = \psi_1^* \phi_1 =
(\Psi , P(\V_p)\Phi )$, so that we can write
\be
(\Psi ,\Phi ) &=& \sum_n (\Psi ,P(\V_p)\Psi_n)(P(\V_p)\Psi_n ,
\Phi )  \cr
&=& \sum_n (\Psi , \Psi_{pn} )(\Psi_{pn} , \Phi ) ~,
\label{4.8}
\en
with a complete set of states $\{\Psi_{pn}\}$ in the Hilbert space
$\V_p$. Writing symbolically $\Psi_{\H n} = \Psi_{pn} +
imQ $, we have $(\Psi ,\Psi_{\H n} ) = (\Psi ,\Psi_{pn})$
for $\Psi \in \H $, and hence obtain the decomposition
\be
(\Psi ,\Phi ) = \sum_n (\Psi , \Psi_{\H n} )
(\Psi_{\H n } , \Phi ) ~,
\label{4.9}
\en
with $\Psi ,\Phi \in \H$. The expression (\ref{4.9})
is manifestly Lorentz
invariant, even though the projection $P(\V_p)$ by itself is not
invariant.
In the full state space $\V$ of indefinite metric, Lorentz
transformations are realized by unitary mappings $U$ with
$U^\dagger = \C U^* \C$. They are BRST-invariant, and
consequently of the form (\ref{4.10}) in our matrix
representation. It is then easy to see that only $U_{11}$
appears in the transformation of physical quantities.
Transformations $U$ with $U_{11} = 1$ are equivalence
transformations which do not change physical matrix elements.
Unphysical states, written as vectors like in Eq.(\ref{4.3} ),
may well have a component in $\V_p$, but we can
always find an equivalence transformation which removes this
component, because $\psi_3 \neq 0$ for these states.

As we have seen in the previous section, we can obtain
hadronic amplitudes as Fourier Transforms of matrix elements
involving only BRST-invariant, local hadronic fields and
hadron states. All spectral conditions result from
decompositions of these products with respect to
intermediate states, which are eigenstates of the energy
momentum operator.
A BRST-invariant operator $O$ commutes with
$Q$, and leaves the subspace $kerQ$ invariant.
In our matrix notation, it is of
the form
\be
O =  \left (
\begin{array}{clcr}
O_{11} & 0      & O_{13}  \\
O_{21} & O_{22} & O_{23}  \\
0      & 0      & O_{33}
\end{array}
\right) ~,
\label{4.10}
\en
with $O_{22} q = q O_{33} $, where $q$ is defined
in Eq.(\ref{4.6}). Since $O\psi \in \H$ if $\Psi \in \H$,
we can use Eq.(\ref{4.9}) to write decompositions
of the form
\be
(\Psi , XY\Phi ) = \sum_n (\Psi , X\Psi_{\H n} )
(\Psi_{\H n} , Y\Phi ) ~,
\label {4.11}
\en
where $\Psi ,\Phi \in \H $ and $X, Y$
are BRST-invariant operators (fields).
We see again, that only physical states appear in the
decomposition.

Eq.(\ref{4.11}) is generic for all spectral decompositions used
in the derivation of analytic properties of physical amplitudes.
It shows that these spectral conditions involve only hadrons,
and it guarantees the unitarity of the S-matrix
\cite{KOJ} in the
physical (hadronic) state space $\H$. The described features
of hadronic amplitudes are, of course, a direct consequence
of our assumption of confinement for transvers gluons and
quarks.

With the local hadronic operator and hadronic spectral
conditions, we have reached the conclusion, that the
derivation of analytic properties, and of dispersion
representations for gauge theories with confinement, can proceed
along the same lines as in the old hadron field theory.
The starting point are Fourier transforms of matrix elements
of retarded and advanced products of the BRST-invariant,
composite, local hadron fields.

However, one important aspect remains to be discussed: the
question of anomalous thresholds or structure singularities,
which will be considered in the following section.
\vskip0.6truein
\centerline {\bf V. ANOMALOUS THRESHOLDS}
\vskip0.2truein

In the literature, anomalous thresholds are often considered
in connection with appropriate Feynman graphs
\cite{KSW, NAM, RHF}. However, they
can be understood, completely independent of perturbation
methods, on the basis of analyticity and unitarity
\cite{RHF}. Within
the framework of theories with confinement, it is essential
to have a nonperturbative approach.

Anomalous thresholds are branch points which appear in a
given channel of an amplitude. They are not directly related
to the possible intermediate states in this channel, which
introduce only ``absorptive'' singularities. They are rather
``structure singularities", which describe effects due to the
possibility that a given particle can be considered as a
composite system of other particles. They appear in the
physical sheet of the amplitude, in the channel considered,
only if a loosely bound composite system is involved. Otherwise,
they remain in a secondary Riemann sheet.

In the following we briefly show that anomalous thresholds,
in a given channel of an amplitude, are due to ordinary
(absorptive) thresholds in crossed channels of other
amplitudes, which are related to the one under consideration
via unitarity. Since we have seen before that hadronic
amplitudes have only absorption thresholds related to hadron
states, it follows that the only anomalous thresholds
which appear are
due to the structure of hadrons as composite systems of
hadronic constituents. There are no such thresholds
associated with the quark-gluon structure of hadrons, even
for loosely bound composite systems with quarks as
constituents.

In order to study
the emergence of anomalous thresholds on the physical sheet
of an hadronic amplitude, we consider a form factor as
an example. We ignore all inessential complications and use
the structure function $W(s)$, $s = k^2$ of a deuteron-like
particle with variable mass, as indicated in Fig.1.
For $x < 2m_N ^2$,  where $x = (mass)^2 \leq m_D^2 $ , the
function $W(s)$ has branch points on the right hand,
real $k^2$-axis, starting with those due to pion
intermediate states at $s_\pi $. However, we
concentrate on the $N\bar{N}$-threshold at $s = 4m_N ^2$.
The discontinuity due to this threshold alone is
\be
Im W_{N\bar{N}} (s + i0) = \rho (s + i0 )
G(s + i0)V^{II} (s + i0) ~,
\label{5.1}
\en
\be
\rho (s) = [(s - 4m_N^2)s^{-1}]^{1/2}  ~.
\label{5.1a}
\en
Here $G(s)$ is the appropriate partial wave projective of
the amplitude $G(s,t)$ pictured in Fig.2a. We consider S-wave
projections for simplicity. Furthermore $V^{II} (s + i0) =
V^* (s + i0)$, where $V(s)$ is the nucleon form factor,
with branch points analogous to those of $W(s)$.
The continuation of $V(s)$ into sheet $II$ of the
$N\bar{N}$-threshold is given by \cite{RHF}
\be
V^{II} (s) = \frac{V(s)}{1 + 2i\rho (s) F(s)}  ~,
\label{5.2}
\en
where $F(s)$ is the partial wave projection of the
scattering amplitude $N\bar{N} \rightarrow N\bar{N}$ in
the s-channel. With Eqs. (\ref{5.1}) and (\ref{5.2}),
we get for the continuation of $W(s)$ through the $N\bar{N}$
cut:
\be
W^{II} (s) = W(s) - 2i\rho (s)\frac{G(s)V(s)}
{1 + 2i\rho (s) F(s)} ~.
\label{5.3}
\en
While $W(s)$ has no left-hand branch lines, $W^{II} (s)$
does, due to left-hand cuts of $G(s)$ and $F(s)$. For our
purpose, the important left-hand cut is the one of $G(s)$,
which is caused by the pole term at $t = m_N^2$ ,
as illustrated in Fig.2b :
\be
G(s,t) = \frac{\Gamma^2 (x)}{m_N^2 - t} + \cdot \cdot \cdot ~.
\label{5.4}
\en
The branch point is due to the end point at $cos \theta = -1 $
in the partial-wave projection, with $t = t(s, cos \theta ) $.
It is located at $s = g(x)$, where
\be
g(x) = 4x \left( 1 - \frac{x}{4m_N^2} \right) ~.
\label{5.5}
\en
For $0 < x < 4m_N^2 $, we have $g(x) < 4m_N^2$,
with the maximum at $g(2m_N^2 ) = 4m_N^2$. The branch
point in sheet II at
$s = g(x)$ is pictured in Fig.3 for $x < 2m_N^2$.

Let us now increase $x$ to $x = 2m_N^2$ and above. The position
$g(x)$ of the branch point moves to $4m_N^2$ at $x = 2m_N^2$,
and then decreases again. Giving $x$ an imaginary part, we get
\be
g(x + iy) = \left( g(x) + \frac{y^2}{m_N^2} \right)
- 2i\frac{y}{m_N^2} (x - 2m_N^2) ~,
\label{5.6}
\en
and we see that $g(x + iy)$ encircles the branch point
$s = 4m_N^2$ of $W(s)$,
moving thereby into the first and ``physical'' sheet
of the Riemann surface of this function. There it becomes an
anomalous threshold. The situation is illustrated
in Fig.4, where the meson cuts have been omitted.
For sufficiently large values of $x$,
this branch point can move well below the lowest absorption
threshold $s_\pi$. Writing $m_D = 2m_N - B$, we get,
for small values of the binding energy $B$,
\be
g(m_D^2) = 4m_D^2 \left( 1 - \frac{m_D^2}{4m_N^2} \right )
\approx 16m_N B ~,
\label{5.7}
\en
which can give a very long maximal range of the distribution
in configuration space, just as expected from the Schr\"{o}dinger
wave function.

There are other anomalous thresholds associated with the
$N\bar{N}$ branch point of $W(s)$. For instance, there are those
due to the probability distribution of the proton in a
deuteron, considered as a composite system of two nucleons and
a limited number of pions. Their position can easily be
calculated exactly. For small values of $B$, we get
$g_1 (m_D^2) \approx 16m_N (B + m_\pi ) $, if
one pion is added.
Anomalous thresholds can also come out of higher absorptive
branch points in the $s$-channel of the form factor $W$.

Finally, we remark that the above considerations can be
generalized to other amplitudes. Essentially only kinematics,
crossing, and some analytic properties are needed.
In all cases the anomalous
thresholds are related to ordinary, absorptive thresholds in
other amplitudes, which appear in the continuation into
secondary Riemann sheets \cite{RHF}.

As we have pointed out before, due to the fact that anomalous
thresholds are indirectly related to absorptive
thresholds, there are {\it no} such singularities which are
associated with the quark-gluon structure of hadrons,
since there are no absorptive thresholds related to this
structure. However, for hadrons which may be considered
as loosely bound systems of heavy quarks,
we can get a large mean square radius on the basis of
appropriate weight functions along {\it hadronic} cuts
\cite{OPN, MEN}, even
though they may be much higher in mass, and also as a consequence
of possible {\it hadronic} anomalous thresholds.

\vskip0.6truein
\centerline{\bf VI. COLORED APMLITUDES}
\vskip0.2truein
Having discussed only hadronic amplitudes describing
observable consequences of the theory, we would like to add
here some remarks about the analytic structure and the
singularities of general Green's functions with colored
channels. In particular, we will show that these colored
amplitudes must have singularities at finite points, which
can be associated with confined states in $\V$ like quarks and
gluons \cite{WZS, OWG}.
Even though quarks and transverse gluons are
confined, we can have asymptotic states associated with
these excitations, as well as corresponding poles in colored
Green's functions. In our formulation of confinement, all
colored states form quartet representations of BRST-algebra,
and hence are not elements of physical space $\H$, which
contains only singlets.

As an example for colored amplitudes, we consider the
gluon propagator, which has been studied extensively. The
structure function is defined as a Fourier Transform by
\begin{eqnarray}
&~&\int dx e^{ikx}   \langle 0 | T A^{\mu \nu}_{a}(x)
A^{\varrho\sigma}_{b}(0) |0\rangle~~ =~~
- i \delta_{ab} D (k^2 + i0)\cr
&~&~~~~~~~\times\left( k^\mu k^\varrho g^{\nu \sigma} -
k^\mu k^\sigma g^{\nu \varrho}
+ k^\nu k^\sigma g^{\mu\varrho} - k^\nu k^\varrho g^{\mu\sigma}\right)
\label{6.1}
\end{eqnarray}
with $A^{\mu \nu}\equiv\partial^\mu A^\nu - \partial^\nu A^\mu$.
As before, we consider the linear, covariant gauges
defined in Eq.(\ref{2.1}). In the state space $\V$ with indefinite
metric, we write the spectral condition in the form
 \cite{STR}
\be
\int d^4 a e^{-ip\cdot a} (\Psi , U(a) \Phi ) = 0 ~,
\label{6.2}
\en
for values of $p$ {\it outside} of $ \overline{W^+} =
\{ p: p^0 \geq 0 , p^2 \geq 0 ; p \in R^4 \}$ , and for
all $\Psi , \Phi \in \V $.

Lorentz covariance and spectral condition are sufficient to
show that $D(k^2 + i0)$ is the boundary value of an analytic function
$D(k^2)$, which is regular in the cut $k^2$-plane, with a cuts
along the positive real axis only. It is then an essential question
to obtain the asymptotic behavior for $k^2
\rightarrow \infty$ in all
directions of the complex $k^2$-plane. In view of the
asymptotic freedom of the theory, the asymptotic terms can
be obtained with the help of renormalization group methods.
In using the renormalization group, an assumption is made,
which we have not used so far. We require that the general
amplitude connects with the perturbative expression
for $g^2 \rightarrow +0 $,
where $g$ is the gauge coupling parameter. The connection is
needed only for the leading term.
With this assumption, we find for $k^2 \rightarrow \infty $
in all directions \cite{OWG}:
\begin{eqnarray}
-k^2 D (k^2,\kappa^2,g,\alpha)&\simeq& \frac{\alpha}{\alpha_0}
 + C_R (g^2, \alpha) \left(-\beta_0 \ln
\frac{k^2}{\kappa^2}\right)^{-\gamma_{00}/\beta_0} + \cdots ~.
\label{6.3}
\end{eqnarray}
The corresponding asymptotic terms for the discontinuity along the
positive, real $k^2$--axis are then given by
\begin{eqnarray}
-k^2 \rho (k^2,\kappa^2,g,\alpha) & \simeq& \frac{\gamma_{0 0}}{\beta_0}
C_R (g^2,\alpha) \left(-\beta_0 \ln \frac{k^2}{\vert
\kappa^2\vert}\right)^{-\gamma_{0 0}/\beta_0 - 1} + \cdots ~.
\label{6.4}
\end{eqnarray}
In these relations, we have used the following definitions:
The anomalous dimension of the gauge field is given by
\be
\gamma (g^2 ,\alpha ) = (\gamma_{00} + \alpha \gamma_{01}) g^2 +
\cdots ~
\label{6.5}
\en
for $g^2 \rightarrow +0$, and for the renormalization
group function we write, in the same limit,
\be
\beta (g^2) = \beta_0 g^4 + \beta_1 g^6 + \cdots ~.
\label{6.6}
\en
Furthermore, we use the notation
\be
\alpha_0  =  -\gamma_{00} /\gamma_{01} ~.
\label{6,7}
\en
For QCD, we have
\be
\frac{\gamma_{00}}{\beta_0}
= \frac{\frac{13}{2}-\frac{2}{3}N_F}{11-\frac{2}{3} N_F} ~,
{}~~~~ \gamma_{01} = (16\pi^2)^{-1} \frac{3}{4} ~,
\label{6.8}
\en
where $N_F$ is the number of quark flavours.
We assume $\beta_0 < 0$ corresponding to asymptotic freedom.
Consequently, the exponent $\gamma_{00}/\beta_0$
in Eqs.(\ref{6.3}) and (\ref{6.4})
varies from $13/22$ for $N_F = 0$ to 1/10 for $N_F = 9$, and from
$-1/16$ for $N_F = 10$ to $-15/2$ for $N_F = 16$.  We have $0 <
\gamma_{00}/\beta_0 < 1$ for $N_F\le 9$ and
$\gamma_{00}/\beta_0 < 0$ for $10 \le N_F \le 16$;
 for $\gamma_{00}/\beta_0 = 1$, our relations
(\ref{6.3}) and (\ref{6.4}) would require modifications.

The parameter $\kappa^2 < 0$ is a normalization point. We generally
chose to normalize $D$ so that
\be
-k^2 D (k^2,\ka^2,g,\alpha) ~= ~1
{}~~~~ \hbox{for} ~~~~ k^2 = \kappa^2~.
\label{6.9}
\en
With this normalization, the coefficient
$C_R(g^2 , \alpha )$ for $\alpha = 0$ is given by
\cite{WZS, OWG}
\be
C_R (g^2,0) &=& (g^2)^{-\gamma_{00}/\beta_0} \hbox{exp}\int^0_{g^2}
dx  \tau_0 (x), \cr
\tau_0 (x) &\equiv& \frac {\gamma (x,0)}{\beta (x)} -
\frac {\gamma_{00}}{\beta_0 x }~,
\label{6.10}
\en
and hence $C_R(g^2 , 0) > 0$.
Certainly $C_R (g^2 , \alpha)$ is not identically
zero. If there should be zero surfaces, a term proportional to
$(-\beta_0 \ln \frac{k^2}{\kappa ^2} )^{-1} $ becomes
relevant in Eq.(\ref{6.4}).

The remarkable property of the asymptotic terms in Eqs.
(\ref{6.3}) and (\ref{6.4}) is their gauge independence
except for the coefficients.
Furthermore, their functional form is
determined by one loop expressions.

{}From the asymptotic limit (\ref{6.3}), we see that
$D(k^2)$ vanishes for $k^2 \rightarrow \infty $
in all directions of the complex $k^2$-plane. Hence
it cannot be a nontrivial entire function, at least for
$0 < g < g_{\infty} $, where $g_{\infty }$
is a possible first non-integrable singularity of
$\beta^{-1} (g^2)$. There must be singularities on the
positive real $k^2$-axis , and it is natural that these are
associated with confirmed, unphysical states. Similar
arguments can be given for the structure functions of the
quark propagator.

We can write an unsubtracted dispersion representation for
$D(k^2 )$ :
\be
 D(k^2,\ka^2,g,\alpha) &=& \int^\infty_{-0} dk'^{2}  \frac{\rho (k'^{2},
 \ka^2, g, \alpha)}{k'^{2} - k^2} ~,
\label{6.11}
 \en
 and even a dipole representation exists
 \begin{eqnarray}
 D (k^2,\ka^2,g,\alpha) &=& \int^\infty_{-0} dk'^{2} \frac{\sigma
 (k'^{2},\ka^2,g,\alpha)}{(k'^{2} - k^2)^2},\cr
 \sigma (k^2,\ka^2,g,\alpha) &=& \int^{k^2}_{-0} dk'^{2} \rho
 (k'^{2},\ka^2,g,\alpha).
 \label{6.12}
 \end{eqnarray}
For $\alpha = 0$, the dipole representation has been used in
order to give arguments for an approximately linear
quark-antiquark potential under the condition
$\gamma_{00}/\beta_0 > 0 $, where $\sigma (\infty ) = 0 $,
and $\sigma (k^2) > 0 $ , $\sigma' (k^2) = \rho (k^2) < 0 $
for sufficiently large values of $k^2$
\cite{RLP, NLP}.

Under the restriction $\gamma_{00} / \beta_0 > 0$
($ N_F \leq 9 $ for QCD),
we find that $D(k^2) - \frac{\alpha}{\alpha_0}$
vanishes faster than $k^{-2}$ at infinity,
so that we have the important sum rule \cite{OWG} :
\be
\int_{-0}^{\infty}d k^2 \rho (k^2, \kappa^2, g, \alpha )
{}~~= ~~\frac{\alpha }{\alpha_0 } ~.
\label{6.13}
\en
For $\alpha = 0$, $\gamma_{00}/\beta_0 > 0$, we have a
superconvergence relation \cite{WZS}.
It gives a rather direct connection between short and long
distance properties of the theory, and has been used in
order to give arguments for gluon confinement
\cite{ROC, NIC}.

\vskip0.6truein
\centerline{\bf APPENDIX: REMARKS ABOUT PROOFS}
\vskip0.2truein

We have seen that we can construct local hadronic fields as
BRST-invariant operators in $\V$, and write Fourier
representation of hadronic amplituds in terms of matrix
elements of products of these fields. With BRST-methods, we
define an invariant physical state space $\H$ which, as a
consequence of confinement, contains only hadrons. With the
spectral conditions also referring to hadrons only, we have
the imput required in order to use the old methods for the
derivation of dispersion representations as formulated in
hadronic field theory. For completeness, we give in this
appendix a very brief sketch of the essential ideas of these
proofs, which are often hidden behind technical details.

The {\it Gap Method} \cite{DSF}
is applicable in cases where there is no
unphysical region. Examples are $\pi\pi$-,
$\pi N$- forward and
near-forward scattering, some form factors like $\pi \pi \gamma$,
$\pi NN$ in the $N$-channel, etc.
\cite{GMO, RON, BOT}.
As an example, let us consider $\pi^0 \pi^0 $-
forward scattering. We can write the amplitude as
\be
F(\omega) = \int_0^\infty dr F(\omega, r) ~,
\label{A.1}
\en
with
\be
F(\omega, r) ~ &=& ~ 4\pi \frac{r ~\hbox{sin}(\sqrt{\omega^2 -
 \mu^2})}{\sqrt{\omega^2  - \mu^2}} \cr
&\times& \int_0^\infty dx^0 e^{i\omega x^0}
\langle p | [j(\frac{x}{2}), j(-\frac{x}{2}) ] | p \rangle ~,
\label{A.2}
\en
and $j = (\Box + \mu^2)\phi $.
For fixed $r$, $F(\omega, r)$ is analytic in the upper
half $\omega$-plane, and $Im F(\omega + i0 , r)$ = 0
for $|\omega| < \mu $ due to the spectral
conditions. Ignoring subtraction,
we can write a Hilbert representation
\be
F ( \omega , r) =
\frac{2 \omega}{\pi}\int_{\mu}^\infty d \omega '
\frac{Im F ( \omega '  + i 0 , r)}{{\omega '}^2 - \omega^2}~~.
\label{A.3}
\en
For real $|\omega | > \mu$, we can perform
the r-integration (\ref{A.1}) on
both sides, and get the corresponding dispersion relation for
$F(\omega)$. Although some refinements are required, the method
shows in a very simple way how local commutativity and spectral
conditions lead to a dispersion representation. Pole terms,
like in $\pi N$- scattering, can also be handled by this method
\cite{BOG, SYM, BOT}.

The {\it General Method} is required in the presence of
unphysical regions, like $NN$-scattering \cite{GNO}
(even for $t$=0), $N\overline {N} \gamma $-form factors in the
$N\overline{N} $-channel,
for fixed $t$
amplitudes \cite{GOR}, to obtain $t$-analyticity
(Lehmann ellipses) \cite{LEH}, and for $st$-analyticity
\cite{STU}.
There are many technical details involved in the derivation
of dispersion representations, like continuations in mass
variables, for example, but the main problem
is to construct the largest
region of holomorphy obtainable on the basis of retarded and
advanced functions like
\be
F_{\pm} (K) = \pm\frac{i}{(2\pi)^3} \int d^4 x~ e^{-i K \cdot x}
\theta (\pm x^0)
{}~\langle p' | \left [ j^{\dagger}(\frac{x}{2}), j(-\frac{x}{2} \right ]
| p \rangle~~,
\label{A.4}
\en
with $K = \frac{1}{2} (k + k') $,~~ $k + p = k' + p'$ ~.

Due to local commutativity, the functions
$F_{\pm} (K) $ are analytic in the wedges
\be
W^{\pm } = \{ K : Im K^0 > 0 ~~or < 0,~~(Im K)^2 > 0;
{}~~Re K \in R^4 \} ~.
\label{A.5}
\en
 From the spectral conditions, we find that
$F_{+} (K) = F_{-} (K)$ for $ D \in R^4$, where $D$
is a real domain, and where this equality may be in the
sense of distributions. As a special case of the {\it Edge
of the Wedge Theorem} \cite{BOT},
we can then show that there exists an
analytic function $F(K)$, which coincides with $F_{\pm }(K)$
in the wedges $W^{\pm}$ respectively,
and which is holomorphic in the domain $W\cup N(D) $, with
$W = W^{+} \cup W^{-} $. Here $N(D)$ is
a finite, complex neighborhood of $D$. If we then construct
the {\it Envelope of Holomorphy} $E(W\cup N(D))$,
we get the largest
possible region of analyticity given the assumptions made.
In the original paper \cite{BOT}, a generalized
semitube has been used, for which the envelope was known.,
This method gives boudary points of the envelope
in important
cases. A complete construction of the envelope, using the
continuitys theorem, was given in \cite{BRO}.

Independently, in \cite{BOG}, elaborate
distribution and analytic methods were used in order to get a
subdomain of $E$, directly on the basis of $W^{\pm}$ and $D$.
For the special problem with one four-vector considered here,
one can use methods from the theory of
distributions and differential equations in order to give a
representation of functions which are holomorphic in $E$
\cite{JLD}.

The limitations of the proofs for dispersion representation
are due to the lack of input from unitarity, and often can
be related to conditions for the absence of unphysical
anomalous thresholds. Some improvements are possible using
aspects of two-particle unitarity, but in general
multiparticle unitarity and analytic properties of
multiparticle amplitudes are required for further
enlargements of the domain of holomorphy.

For any fixed $t < 0$, and for arbitrary binary reactions, it
can be shown that the amplitude is
holomorphic outside of a large circle in the cut $s$-plane,
so that one can always prove crossing relations \cite{BEG}.

As is evident from the preceding discussion, the
interesting proposal of double dispersion relations
\cite{MAN} has not been proven. They are compatible with
hadronic perturbation theory in lower orders. Although
it may not be a valid approach in QCD, hadronic
perturbation theory is a useful tool for locating certain
singularities of physical amplitudes.

\vskip0.7truein
\centerline{\bf ACKNOWLEDGEMENTS}
\vskip0.2truein

I am indebted to Gerhard H\"{o}hler for insisting that
I should write an article about the validity of
dispersion relations in QCD. For conversations or remarks,
I would like to thank H.-J. Borchers, G. H\"{o}hler,
J. Kubo, H. Lehmann, K. Nishijima, B. Schroer,
K. Sibold, F. Strocchi and W. Zimmermann. It is a pleasure
thank Wolfhart Zimmermann, and the
Theory Group of the Max Planck Institut
f\"{u}r Physik, Werner Heisenberg Institut, for their
kind hospitality in M\"{u}nchen.

This work has been supported in part by the
National Science Foundation, grant PHY 91-23780.
\newpage
\vskip0.7truein

\newpage

\vskip0.6truein
\centerline{\bf Figure Captions}
\vskip1.0truein

\newcounter{fig}
\begin{list}
{\bf Fig. \arabic{fig}.}{\usecounter{fig}
  \labelwidth1.6cm \leftmargin2.5cm\labelsep0.4cm \rightmargin1cm
  \parsep1.5ex plus0.2ex minus0.1ex \itemsep2.5ex plus0.2ex }
\item   Vertex Function $W(s)$.
\item   Inelastic amplitude
        $G(s,t)$ (a) and relevant pole term (b) in the $t$-channel.
\item   Branch points of $W(s)$ and $W^{II} (s) $. The continuation
        is with respect to the $N\overline {N}$-threshold at
        $s = 4m_N^2$.
\item   Anomalous threshold of $W(s)$ at $s = g(x)$ for
        $x > 2m_N^2 $. The branch line runs from $g(x)$
        to $4m_N^2 $ in sheet I (physical sheet), and
        then from $4m_N^2$ to $-\infty $ in sheet II
        (dotted line). The meson branch lines starting
        at $s_\pi < 4m_N^2  $ have not been drawn.
\end{list}


\begin{thebibliography}{99}
\bibitem{BOT}
{H. J. Bremermann, R. Oehme and J. G. Taylor,
Phys. Rev. {\bf 109} (1958) 2178.}
\bibitem{BRO}
{J. Bros, A. Messiah and R. Stora,
Journ. Math. Phys. {\bf 2} (1961) 639.}
\bibitem{MAT}
{A. Martin, in {\it Lecture Notes in Physics}
No. {\bf 3} (Springer Verlag, Berlin, 1969);
G. Sommer, Fortschritte der Physik, {\bf 18} (1970) 577;
and papers quoted in these articles. }
\bibitem{JLD}
{R. Jost and H. Lehmann, Nuovo Cimento  {\bf 5},
1958 (1957);
F. J. Dyson, Phys. Rev. {\bf 110} (1958) 1460. }
\bibitem{BOG}
{N. N. Bogoliubov, B. V. Medvedev and M. V. Polivanov,
{\it Voprossy Teorii Dispersionnykh Sootnoshenii}
(Fitmatgiz, Moscow, 1958);
N. N. Bogoliubov and D. V. Shirkov, {\it Introduction to
the Theory of Quantized Fields} (Interscience,
New York, 1959).}
\bibitem{DSF}
{R. Oehme, Nuovo Cimento {\bf 10} (1956) 1316; \\
R. Oehme, in {\it Quanta}, edited by P. Freund,
C. Goebel and Y. Nambu (University of Chicago Press,
Chicago, 1970) pp. 309-337. }
\bibitem{SYM}
{K. Symanzik, Phys. Rev. {\bf 100} (1957) 743. }
\bibitem{LEH}
{H. Lehmann, Suppl. Nuovo Cimento, {\bf 14} (1959) 1;~
Nuovo Cimento {\bf 10} (1958) 1460. }
\bibitem{OTG}
{R. Oehme and J. G. Taylor, Phys. Rev. {\bf 113} (1959) 371.}
\bibitem{OPN}
{R. Oehme, Mod. Phys. Lett. {\bf 8}, 1533 (1993);
$\pi N$-Newsletter No. {\bf 7} (1992) 1,
Fermi Institute Report EFI 92-17,  (unpublished). }
\bibitem{WZC}
{W. Zimmermann, Nuovo Cimneto {\bf 10} (1958) 596;
K. Nishijima, Phys. Rev. {\bf 111} (1958) 995.}
\bibitem{WLS}
{K. Wilson, Phys. Rev. {\bf 179} (1968) 1499;
K. Wilson and W. Zimmermann, Comm. Math. Phys. {\bf 24} (1972) 87;
W. Zimmermann, in {\it 1970 Brandeis Lectures}, edited by
S. Deser, M. Grisaru and H. Pendleton (MIT Press,
Cambridge, 1971) pp. 395-591;
W. Zimmermann, in {\it Wandering in the Fields}, edited by
K. Kawarbayashi and A. Ukawa (World Scientific,
Singapore, 1987) pp. 61-80.}
\bibitem{BRS}
{C. Becchi, A. Rouet and R. Stora, Ann. Phys. (N.Y.)
{\bf 98} (1976) 287; I. V. Tyutin, Lebedev Report
 No. FIAN 39 (1975) (unpublished). }
\bibitem{SPI}
{M. Spiegelglas, Nuc. Phys. {\bf B283} (1987) 205. }
\bibitem{KOJ}
{T. Kugo and I. Ojima, Prog. Theor. Phys. Suppl.
{\bf 66} (1979) 1;
N. Nakanishi, Prog. Theor. Phys. {\bf 62} (1979) 1396;
K. Nishijima, Nucl. Phys. {\bf B238} (1984) 601;
I. B. Frenkel, H. Garland and G. J. Zuckerman,
Proc. Nat. Acad. Sci. USA, {\bf 83} (1986) 8442;
R. Oehme, Mod. Phys. Lett. {\bf A6} (1991) 3427;
N. Nakanishi and I. Ojima, {\it Covariant Operator
Formalism of Gauge Theories and Quantum Gravity}
(World Scientific, Singapore, 1990). }
\bibitem{ROC}
{R. Oehme, Phys. Rev. {\bf D42} (1990) 4209; ~
Phys. Lett. {\bf B155} (1987) 60. }
\bibitem{NIC}
{K. Nishijima, Prog. Theor. Phys. {\bf 75} (1986) 22;
K. Nishijima and Y. Okada, ibid. {\bf 72} (1984) 254;
K. Nishijima in {\it Symmetry in Nature}, Festschrift
for Luigi A. Radicati di Brozolo (Scuola Normale Superiore,
Pisa, 1989) pp. 627-655. }
\bibitem{RAN}
{R. Oehme, Phys. Rev {\bf 111} (1958) 143;
Nuovo Cimento {\bf 13} (1959) 778. }
\bibitem{NAM}
{Y. Nambu, Nuovo Cimento {\bf 9} (1958) 610.}
\bibitem{KSW}
{R. Karplus, C. M. Sommerfield and F. H. Wichmann,
Phys. Rev. {\bf 111} (1958) 1187;
L. D. Landau, Nucl. Phys. {\bf B13} (1959) 181;
R. E. Cutkosky, J. Math. Phys. {\bf 1} (1960) 429. }
\bibitem{RHF}
{R. Oehme, in {\it Werner Heisenberg und die Physik
unserer Zeit}, edited by F. Bopp (Vieweg, Braunschweig,
1961) pp. 240-259;
Phys. Rev. {\bf 121} (1961) 1840. }
\bibitem{WZS}
{R. Oehme and W. Zimmermann, Phys. Rev. {\bf D21} (1980) 474, 1661. }
\bibitem{OWG}
{R. Oehme and W. Xu, Phys. Lett. {\bf B333}, (1994) 172. }
\bibitem{SKU}
{T. Kugo and S. Uehara, Prog. Theor. Phys. {\bf 64} (1980) 1395. }
\bibitem{SKO}
{Kugo, Ojima \cite{KOJ}.}
\bibitem{SNA}
{Nakanishi  \cite{KOJ} .}
\bibitem{ROS}
{R. Oehme, Phys. Lett. {\bf B252} (1990) 641.}
\bibitem{WZK}
{Zimmermann \cite{WLS} }
\bibitem{LSZ}
{H. Lehmann, K. Symanzik and W. Zimmermann,
Nuovo Cimento {\bf 1} (1955) 425;  {\bf 6} (1957) 319. }
\bibitem{BOC}
{H.-J. Borchers, Nuovo Cimento {\bf 15} (1960) 784. }
\bibitem{MEN}
{R. L. Jaffe and P. F. Mende, Nucl. Phys. {\bf B369} (1992) 189. }
\bibitem{STR}
{F. Strocchi, Comm. Math. Phys. {\bf 56} (1978) 57;
Phys. Rev. {\bf D17} (1978) 2010. }
\bibitem{RLP}
{R. Oehme, Phys. Lett. {\bf B232} (1989) 489. }
\bibitem{NLP}
{K. Nishijima, Prog. Theor. Phys. {\bf 77} (1987) 1053. }
\bibitem{GMO}
{M. L. Goldberger, H. Miyazawa and R. Oehme, Phys Rev.
{\bf 99} (1956) 986. }
\bibitem{RON}
{R. Oehme, Phys. Rev. {\bf 100} (1955) 1503;
{\bf 101} (1956) 1174. }
\bibitem{GNO}
{M. L. Goldberger, Y. Nambu and R. Oehme, Ann. Phys.
(N.Y.)  {\bf 2} (1956) 226. }
\bibitem{GOR}
{M. L. Goldberger, Y. Nambu and R. Oehme, reported in
{\it Proceedings of the Sixth Annual Rochester Conference}
(Interscience, New York, 1956) pp. 1-7 ;
G. F. Chew, M. L. Goldberger, F. E. Low and Y. Nambu,
Phys. Rev. {\bf 106} (1957) 1337. }
\bibitem{STU}
{S. Mandelstam, Nuovo Cimento {\bf 15} (1964) 658;  \\
H. Lehmann, Comm. Math. Phys. {\bf 2} (1966) 375. }
\bibitem{BEG}
{J. Bros, H. Epstein and V. Glaser, Comm. Math. Phys.
{\bf 1} (1965) 240. }
\bibitem{MAN}
{S. Mandelstam, Phys. Rev. {\bf 112} (1958) 1344. }





\end{thebibliography}
\end{document}